\documentclass[prl,twocolumn,showpacs]{revtex4}
\usepackage{graphicx}
\usepackage{amsmath}
\usepackage{amssymb}

\begin{document}
\title{Particle-Hole Symmetry Breaking and the 5/2 Fractional Quantum Hall Effect}

\author{Hao Wang$^{1}$, D. N. Sheng$^1$, and F. D. M. Haldane$^2$}
\affiliation{$^{1}$Department of Physics and Astronomy, California
State University, Northridge, California 91330, USA \\ $^2$
Department of Physics, Princeton University, New Jersey 08544, USA}


\begin{abstract}
We report on the study of the fractional quantum Hall effect at the
filling factor 5/2 using exact diagonalization method with torus
geometry. The particle-hole symmetry breaking effect is considered
using an additional three-body interaction. Both Pfaffian and
anti-Pfaffian states can be the ground state depending on the sign
of the three-body interaction. The results of the low-energy
spectrum, the wave function overlap, and the particle-hole parity
evolution, have shown the clear evidence of a direct sharp
transition (possibly first-order) from the Pfaffian to the
anti-Pfaffian state at the Coulomb point. A quantum phase diagram is
established, where one finds  further transitions from the
Pfaffian or anti-Pfaffian state to the nearby compressible phases
induced by a change of the pseudopotential.
\end{abstract}

\maketitle

The fractional quantum Hall effect (FQHE) at the filling factor $\nu
=5/2$ has recently drawn intensive attentions in theoretical and
experimental
studies~\cite{exp,pf,pf1,pf2,pf3,3b,3b1,3b2,appl,apf0,apf,pete}. The
Moore-Read Pfaffian (Pf) state has been proposed as a successful
candidate to describe the $5/2$ FQHE \cite{pf,pf1,pf2}. The Pf state
as a non-Abelian topological phase suggests a potential application
towards the quantum computing \cite{appl}. However, the Pf state
breaks particle-hole (PH) symmetry and its PH conjugate state, known
as the anti-Pfaffian (APf) state, has been suggested to be another
candidate state \cite{apf0,apf}. In the limit of vanishing
Landau-level (LL) mixing, the Pf and APf model wave functions have
the same energy for the $5/2$ FQHE system with the pure Coulomb
interaction as the system has the PH symmetry. Numerically the
ground state (GS) of such a system appears as a superposition of the
Pf and APf states\cite{pf2,pf3}. However, the PH symmetry of the
Hamiltonian in real systems can be broken by many factors, such as
LL mixing. This raises a new challenge regarding the nature of the
GS realized in the experimental system\cite{apf0, apf}, which can be
the Pf, APf, or superposition of them. To address this issue, we add
a PH nonsymmetric three-body (3b) interaction together with the
Coulomb interaction  as the model Hamiltonian to reexamine the
nature of the GS and to study the quantum phase transitions (QPT)
between distinct quantum phases.

In this letter, We investigate the low-energy states of the 5/2 FQHE
system using Lanczos method for finite-size systems with up to
$N_e=14$ electrons. The GS of a Coulomb system is found to have a
transition into the Pf or APf state with the turn-on of the 3b
interaction. The Pf and APf states are robust in a range of the 3b
interaction, insensitive to the detailed form of the 3b interaction
as long as the PH symmetry is broken.  The calculations on the GS
wave function overlap, energy and PH parity evolutions show the
strong evidence that the phase transition between the Pf and the APf
state appears to be first-order, occurring exactly at the Coulomb
point. Under an extra short-range pseudopotential\cite{pf2},  Pf and
APf states can have transition to the stripe phase or composite
fermion liquid (CFL) phase.

We consider a two-dimensional electron  system  under a
perpendicular magnetic field. Periodic boundary conditions for
magnetic translational operators are imposed with a quantized flux
$N_{\phi}$ through a rectangular unit cell $\mathbf{L_1} \times
\mathbf{L_2}$. The magnetic length $\ell$ is taken as the unit of
the length and the energy is in units of $e^2/4\pi \epsilon \ell$.
To reduce the size of the Hilbert space, we carry out our
calculation at every pseudomomentum $\mathbf{K}=(K_1,K_2)$
\cite{haldane}, where $K_1(K_2)$ is in unit of $2\pi/L_1 (2\pi/L_2)$
and the even number of electrons is used for $N_e$. The magnetic
field is assumed to be strong enough so that the spin degeneracy of
the LLs is lifted\cite{pf1,pf2,3b2}. One can thus project the system
Hamiltonian into the topmost, half filled, $N=1$ LL \cite{pf2}. The
projected Hamiltonian for the Coulomb interaction has the form:
\begin{eqnarray}
  H_{c}=
    \frac{2}{N_{\phi}} \sum_{i<j}\sum_{\mathbf{q}} e^{-q^2/2} e^{i \mathbf{q} \cdot (\mathbf{r}_i-\mathbf{r}_j)}
    \sum_{m=0}^{\infty} V_{m}L_{m}(q^2),
\end{eqnarray}
where $V_{m}$ is the Haldane's pseudopotential of the Coulomb
interaction at $N=1$ LL and $L_m(x)$ is the Laguerre polynomial. The
momentum $\mathbf{q}$ is taken discrete values suitable for the unit
cell lattice. $\mathbf{r}_i$ is the guiding center coordinate of the
$i$-th electron.

The Pf state on torus can be obtained as the zero-energy GS of a
repulsive 3b potential given by \cite{pf2}:
\begin{eqnarray}
  H_{3}=-\sum_{i<j<k}S_{i,j,k}
  [\nabla_{i}^4 \nabla_{j}^2 \delta^2(\mathbf{r}_i-\mathbf{r}_j)
  \delta^2(\mathbf{r}_j-\mathbf{r}_k)],
\end{eqnarray}
where $S_{i,j,k}$ is a symmetrizer. Besides the  center-of-mass
degeneracy, the Pf state appears at three particular pseudomomenta
of $(0,N_e/2),(N_e/2,0),(N_e/2,N_e/2)$ \cite{pete} as degenerate
ground states. To study the effect of the PH symmetry breaking, we
set up a model Hamiltonian with the form $H=H_c+V_{3b}\cdot H_3$,
where the parameter $V_{3b}$ can change its sign and magnitude.

\begin{figure}
\centerline{\includegraphics [width=3.2 in] {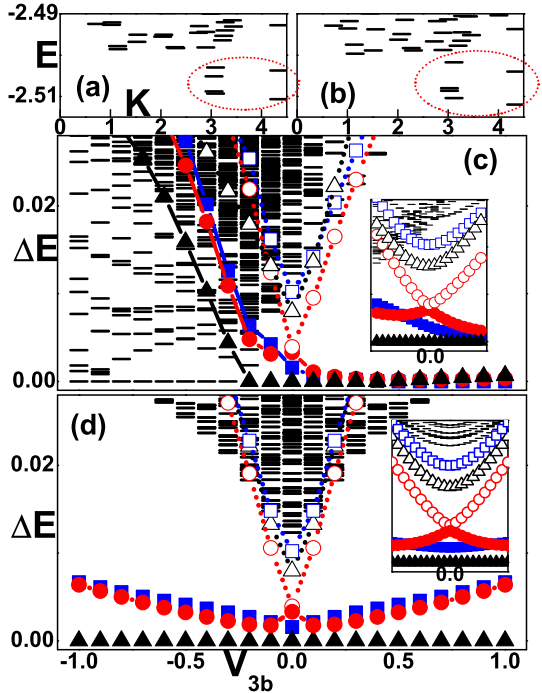}} \caption{(Color
online) Low-lying energies as a function of the pseudomomentum
$\sqrt{K_1^2+K_2^2}$ (in physical units) for a $N_e=12$ pure Coulomb
system with the rectangular unit cell at the aspect ratios of (a)
$L_1/L_2=0.99$, and (b) $L_1/L_2=0.97$, where a pair of
lowest-energy  states at the pseudomomentum (0,6) are nearly
degenerate. The plots of (c) and (d) are low-lying energies
(relative to the globe GS) as a function of $V_{3b}$ for the
$N_e=12$ model system with the additional 3b interaction $H_3$ and
$H^{*}_3$, respectively. The unit cell is the same as the one in
(b). The circle, square and triangle symbols stand for the GSs
(solid symbols) and the first excited states (open symbols) at the
pseudomomenta of (0,6), (6,0) and (6,6), respectively. The insets in
(c) and (d) are the zoom-in spectra for $-0.1<V_{3b}<0.1$.}
\end{figure}

We first present the low-energy spectrum as shown in Fig. 1(a) and
Fig.1 (b) for a pure Coulomb system with $N_e=12$ electrons. The
low-energy sector consists of six states from the three
pseudomomenta of (0,6), (6,0) and (6,6), three local GSs of which
are termed as the GS triplet, separating from the other excited
states. The wave function overlaps with Pf and APf states
demonstrate that these six states are closely related to the
superposition states of Pf and APf. By adjusting the aspect ratio of
the unit cell from $L_1/L_2=0.99$ to 0.97 the lowest-energy pair
states at the pseudomomentum (0,6) change from well separated to
nearly degenerate (see also the inset of Fig. 1(c) and (d)). This is
consistent with the fact that the tunneling between Pf and APf in a
finite-size Coulomb system is sensitive to the specific geometry of
the system. The fine tuning of the aspect ratio reduces the
finite-size effect by reducing the energy gap between the lowest two
states at the particular pseudomomentum.

Fig. 1(c) exhibits the low-lying excitation spectrum as a function
of $V_{3b}$ for the $N_e=12$ model system with the additional 3b
interaction $H_3$. The unit cell is the same as the one described in
Fig. 1(b). At the Coulomb point ($V_{3b}=0$), each GS in the GS
triplet and its first excited state have the opposite PH parity. In
the region of $V_{3b}>0$, with $H_3$ strength getting stronger, the
degeneracy of the GS triplet improves and the energy gap between the
GS triplet and other states increases. In the region of $V_{3b}<0$,
with the 3b strength $|V_{3b}|<0.1$ the near degeneracy of the GS
triplet maintains. With more negative $V_{3b}$, the energy width  of
the lowest triplet increases and these energy levels cross with
higher energy states, indicating a phase transition induced by
$-H_3$.

We further check if the obtained results depend on the precise form
of the 3b interaction. Noting that $H_3$ has the PH symmetric and
anti-symmetric components,  we extrapolate the PH anti-symmetric
component of $H_3$, which has the form of
$H^{*}_3=(H_3-\Tilde{H_3})/2$ with $\Tilde{H_3}$ as the PH conjugate
of $H_3$, as the other type of additional 3b interactions. This way,
$H=H_c+V_{3b}\cdot H^*_3$ has the generic form of both symmetric and
anti-symmetric parts. The obtained low-energy spectrum
 is shown in Fig. 1(d). Due to the PH
anti-symmetry of the $H^{*}_3$, the spectrum is symmetric around the
Coulomb point. Clearly, the three lowest energy states are well
separated from the excited states, indicating the establishing of Pf
or APf states.

\begin{figure}
\centerline{\includegraphics [width=3.2 in] {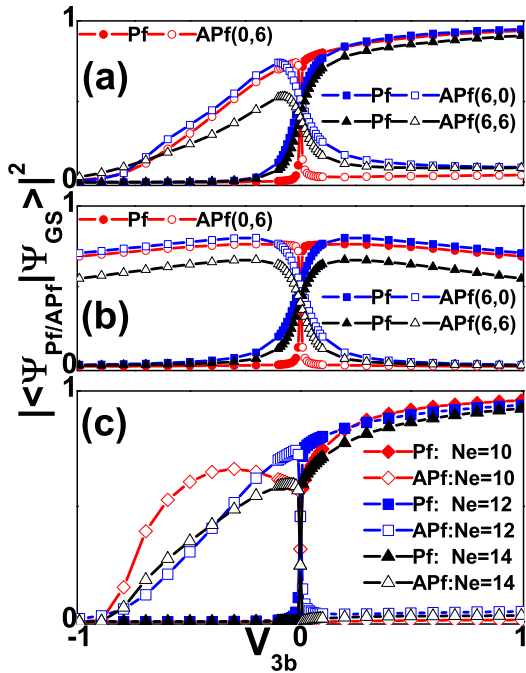}} \caption{(Color
online) The projection of the system GS on the Pf and APf states as
a function of $V_{3b}$. The aspect ratio $L_1/L_2$ are 0.94, 0.97
and 0.70 for the system with $N_e=$10, 12 and 14 electrons,
respectively. The curves with solid (open) symbols stand for the
projection on the Pf (APf) state. States of $N_e=12$ system with the
additional 3b interaction of (a) $H_3$ and (b) $H^{*}_3$ at
pseudomomenta (0,6), (6,0) and (6,6) are represented by the curves
with circle, square and triangle symbols, respectively. (c) States
of $N_e=$10 system with $H_3$ at pseudomomentum (5,5), $N_e=12$ at
(0,6) and $N_e=14$ at (7,7) are represented by the curves with
diamond, square and triangle symbols, respectively.}
\end{figure}

To explore the nature of the GS under the additional 3b interaction,
in Fig. 2(a) we plot the wave function overlap (squared) between the
GS triplet and a Pf or an APf state as a function of $V_{3b}$ for
the $N_e=12$ model system with the 3b interaction $H_3$ and the same
unit cell as in Fig. 1(b). In the positive $V_{3b}$ region, with the
$H_3$ strength growing, the GS projection on the Pf state
monotonously increases towards unitary while the projection on the
APf state decreases and drops to a small value, indicating that the
GS is in the same class as the Pf state. In the negative $V_{3b}$
region, with the strength of $-H_3$ increasing, the projection on
the Pf state quickly decreases and tends to be zero. On the other
hand, the GS projection on the APf state, up to the strength
$|V_{3b}|\sim 0.1$, increases and remains at a finite value between
0.6 and 0.7, indicating that the GS is associated with the APf in
this region. When the strength increases further, the GS projection
on the APf state continuously drops and eventually vanishes. The
results for the model system with $H^{*}_3$ and the same unit cell
have been shown in Fig. 2(b). In the positive $V_{3b}$ region, the
projections on the APf state tend to vanish as the 3b strength grows
while the projections on the Pf state increase to some finite value
between 0.6 and 0.7 with the strength up to 0.1, indicating the GS
is associated with the Pf state. In the negative $V_{3b}$ region,
the GS has large overlap with  the APf phase. The above
characteristics from the wave function overlaps agree with the
spectrum feature shown in Fig. 1(c) and (d). We also notice that the
projection curves of the local GS at the pseudomomentum (0,6), which
has the nearly degenerate first excited state, exhibit the sharpest
transition in a small region of $V_{3b}$ crossing the Coulomb point,
indicating the strongest QPT signal. In the following discussion we
target such particular GS from the GS triplet to investigate the
phase transition between the Pf and the APf state. In Fig. 2(c), the
results of the wave function overlap for different sizes of the
system with the additional 3b interaction $H_3$ have been
demonstrates. For all systems considered, the Pf state dominates on
the $V_{3b}>0$ side while the APf states dominates on the $V_{3b}<0$
side as long as the 3b interaction strength is smaller than $0.1$
and a sharp transition occurs at the Coulomb point.

\begin{figure}
\centerline{\includegraphics [width=3.2 in] {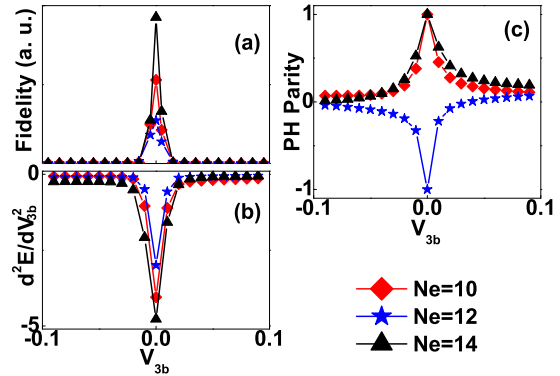}} \caption{(Color
online) Detectors of the QPT for the model system with $H_3$ at
$-0.1<V_{3b}<0.1$ and different $N_e$. (a) GS fidelity function, (b)
second derivative of the GS energy, and (c) GS PH parity as function
of $V_{3b}$. The curves with diamond, star and triangle symbols
represent for the system with $N_e=$10, 12 and 14 electrons,
respectively.}
\end{figure}

The low-lying energy spectrum and the GS wave function projections
provide us with a convincing picture for the existence of a QPT
towards two different states, the Pf and APf states, with the
turn-on of the additional 3b interaction. To gain the further
understanding for this QPT, we study several physical quantities for
a model system near the Coulomb point. In Fig. 3 we show the
evolutions of the fidelity function~\cite{fid2}, energy and PH
parity of the chosen GS for different system sizes when the strength
of the 3b interaction $H_3$ varies in the range $|V_{3b}|<0.1$ with
the increment $\delta V=0.01$. The pseudomomenta of the chosen GSs
and the unit cells for $N_e=10,12,14$ systems are the same as those
in Fig. 2(c). The GS fidelity function, which we use in the plot of
Fig. 3(a) to probe the response of the wave function to the
variation of the parameter $V_{3b}$, has the form of
$F(V_{3b})=-\frac{Ln[|\langle\Psi(V_{3b}-\delta
V)|\Psi(V_{3b}+\delta V)\rangle|^2]}{\delta V^2}$. The GS wave
function is found insensitive to the change of the 3b interaction
when its strength is larger than 0.02. However, within the strength
range $|V_{3b}|<0.02$, the value of the fidelity function abruptly
increases, indicating a QPT occurs. The peak around $V_{3b}=0$
becomes sharper if we improve the degeneracy between the chosen GS
and its first excited state by carefully tuning the parameters of
the unit cell. With the sharp peak located at $V_{3b}=0$ for all
different system sizes, we can identify the Coulomb point as the
transition critical point. We can also trace the GS energy to probe
a QPT. In Fig. 3(b) we plot the second derivative of the GS energy
as a function of $V_{3b}$. For all the system sizes considered, the
curves exhibit the singularity-like behavior at the Coulomb point,
signalizing a first-order QPT. Another useful protocol for us to
understand the QPT crossing $V_{3b}=0$ is the PH parity of the GS as
shown in Fig. 3(c). Within a narrow range of $|V_{3b}|<0.03$, the PH
parity of the GS quickly collapses from the Coulomb point, where it
is either unitary or anti-unitary, confirming the relation between
the QPT and the PH symmetry breaking.

\begin{figure}
\centerline{\includegraphics [width=3.2 in] {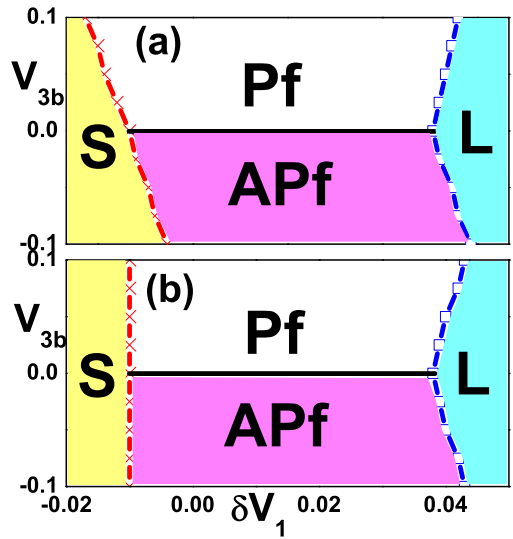}} \caption{(Color
online) Phase diagram of $N_e=12$ system near the Coulomb point with
the 3b interactions of (a) $H_3$ and (b) $H^{*}_3$. The letters of
S, L, Pf and APf are used to mark the phase regions of ($\delta
V_1$,$V_{3b}$) with the stripe-like, composite-fermion-liquid-like,
Pf and APf phase, respectively.}
\end{figure}

In a generic system with the finite LL mixing, one would expect some
effective interaction terms breaking the PH symmetry, which are now
modeled by $H^*_{3}$. One would also expect some effective terms
with the PH symmetry, which may be modeled as a change of the
Haldane's pseudopotential \cite{pf2}. We map out the quantum phase
diagram for the 5/2 FQHE system in the presence of the 3b
interaction and an extra short-range pseudopotential $\delta V_1$.
Fig. 4(a) and (b) show the phase diagram of the $N_e=12$ model
system with the 3b interaction $H_3$ and $H^{*}_3$, respectively. We
only concern the 3b interaction with the strength less than 0.1,
where the obtained phase is most relevant for a Coulomb system with
the weak PH symmetry breaking. Similar to the story in a pure
Coulomb system\cite{pf2}, we find that the model system with the
weak additional 3b interaction will transfer from the Pf-like or
APf-like state to the compressible stripe phase when we reduce the
pseudopotential $\delta V_1$ and to the CFL phase when we increase
$\delta V_1$. Between these two compressible phases, the Pf phase is
found associated with the positive $V_{3b}$ and the APf phase with
the negative $V_{3b}$. For the boundary between the stripe and the
Pf/APf phase, at each $V_{3b}$ the boundary point $\delta V^{S}_1$
locates where the peak value of the structure factor as a function
of $\delta V_1$ drops most dramatically. For the boundary between
the Pf (APf) and CFL phases, at each $V_{3b}$ with a positive
(negative) sign, the boundary point $\delta V^{L}_1$ locates where
the GS projection on the CFL state begins to exceed the one on the
Pf (APf) state with the square unit cell considered. As shown in the
Fig. 4(a), on the $\delta V_1>0$ side both the Pf and the APf phase
tend to expand their phase boundaries with the CFL phase when the 3b
interaction strength grows. On the $\delta V_1<0$ side, the Pf phase
extends its boundary with the stripe phase while the APf phase
shrinks. The nonsymmetric behaviors of the Pf and APf phase result
from the PH symmetric component of $H_3$. The phase diagram of the
system with 3b interaction $H^{*}_3$, as shown in Fig. 4(b), is
symmetric around the phase line $V_{3b}=0$.

In summary, the PH nonsymmetric 3b term modeling a realistic 5/2
FQHE system can bring either the Pf or the APf state as the ground
state depending on its sign. The pure Coulomb system is at the
critical point for a possible first-order transition between these
two states as one changes the sign of the 3b interaction. Our
results suggest that the APf state is indeed a valid
candidate\cite{apf0, apf} for the experimental observed $5/2$ FQHE.

\section{Acknowledgments}
This work is supported by the U.S. NSF grants DMR-0605696 and
DMR-0611562, the DOE grant DE-FG02-06ER46305 (HW, DNS), and NSF
MRSEC program, Grant No. DMR-0819860 (FDMH).
We also thank the KITP for support through the NSF grant PHY05-51164.

\end{document}